\documentclass[a4paper]{jpconf}
\usepackage{graphicx}
\begin{document}
\title{Elliptic flow of $\Lambda$, $\Xi$ and $\Omega$ in 2.76 A TeV Pb+Pb collisions}

\author{Huichao Song,  Fanli Meng, Xianyin Xin, Yu-Xin Liu}

\address{Department of Physics and State Key Laboratory of Nuclear Physics and Technology, Peking
University, Beijing 100871, China}

\address{Collaborative Innovation Center of Quantum Matter, Beijing 100871, China}

\ead{huichaosong@pku.edu.cn}

\begin{abstract}
The elliptic flow $v_2$ for strange and multi-strange bayons in 2.76 A TeV Pb+Pb collisions is investigated with {\tt VISHNU} hybrid model that connects 2+1-d viscous hydrodynamics with a hadron cascade model. It is found that {\tt VISHNU} nicely describes $v_2(p_T)$ data for $\Lambda$, $\Xi$ and $\Omega$ at various centralities. Comparing with the ALICE data, it roughly reproduces the mass-ordering of $v_2$ among $\pi$, K, p, $\Xi$ and $\Omega$ within current statistics, but gives an inverse $v_2$ mass-ordering between p and $\Lambda$.
\end{abstract}

\section{Introduction}

\quad The QGP viscosity is a hot research topic. Viscous hydrodynamic simulations showed that elliptic flow $v_2$ is very sensitive to the QGP shear viscosity to entropy density ratio $\eta/s$. Even the minimum value $1/(4\pi)$ from the AdS/CFT correspondence could lead to 20-25\% suppression of $v_2$~\cite{Song:2007fn,Romatschke:2007mq}. Generally, the QGP viscosity is extracted from the integrated $v_2$ for all charged hadrons, since it is directly related to the fluid momentum anisotropy and varies monotonously with $\eta/s$~\cite{Song:2010mg,Song:2011hk}. The anisotropy flow for common hadrons (pions, kaons and protons) are developed in both QGP and hadronic stages. A precise extraction of the QGP viscosity from these soft hadron data requires a
sophisticated hybrid model that nicely describes the kinetics and decoupling of the hadronic matter. In contrast, hadrons contain multiple strange quarks ($\Xi$, $\Omega$ and etc.) are predicted to decouple earlier from the system due to their smaller hadronic cross sections~\cite{Adams:2003fy,Estienne:2004di}. They may directly probe the QGP phase and can be used to test the QGP viscosity extracted from the elliptic flow of common hadrons~\cite{Song:2012ua}.

Recently, the ALICE collaboration measured the multiplicity, $p_T$ spectra and elliptic flow for strange and multi-strange hadrons~\cite{ABELEV:2013zaa,Zhou:2013uwa}, so this is the right time to perform a careful theoretical investigation of these data. In this article, we will briefly report {\tt VISHNU} calculations of the elliptic flow for $\Lambda$, $\Xi$ and $\Omega$ hyperons in 2.76 A TeV Pb+Pb collisions.  Detailed investigation of these strange and multi-strange hadrons can be found in the incoming article~\cite{Song2013-2}. The {\tt VISHNU} predictions for the elliptic flow of $\phi$ meson are documented in Ref~\cite{Song2013}.

\section{Set-ups}
\quad The theoretical tool used here is {\tt VISHNU} hybrid model~\cite{Song:2010aq}. It connects 2+1-d hydrodynamics for the viscous QGP fluid expansion to a hadron cascade model for the kinetic evolution of the hadronic matter. For simplicity, we neglect the net baryon density, heat flow and bulk viscosity. The equation of state (EOS) s95p-PCE used for the hydrodynamic evolution is constructed from recent lattice data~\cite{Huovinen:2009yb}.
The hyper-surface switching between hydrodynamics and the hadron cascade is constructed by a switching temperature at $165 \texttt{ MeV}$, which is close to the QCD phase transition and chemical freeze-out temperature at top RHIC energy. Following Ref~\cite{Song:2011qa}, we use MC-KLN initial conditions. The initial entropy density profiles for selected centralities are generated from MC-KLN model through averaging over a large number of fluctuating entropy density distributions with recentering and aligning the reaction plane. The QGP specific shear viscosity $(\eta/s)_{QGP}$ is set to 0.16, which gives a nice description of $v_2(p_T)$ for pion, kaon and protons at the LHC~\cite{Song2013}. For $(\eta/s)_{QGP}=0.16$, the starting time $\tau_0$ is 0.9 fm/c, which is obtained from fitting the slope of the $p_T$ spectra for all charged hadrons below 2 GeV~\cite{Song:2011qa}. With these inputs and parameters fixed, we predict the elliptic flow for strange and multi-strange bayons in 2.76 A TeV Pb+Pb collisions.
\begin{figure*}[t]
\includegraphics[width=0.95\linewidth,height=9cm,clip=]{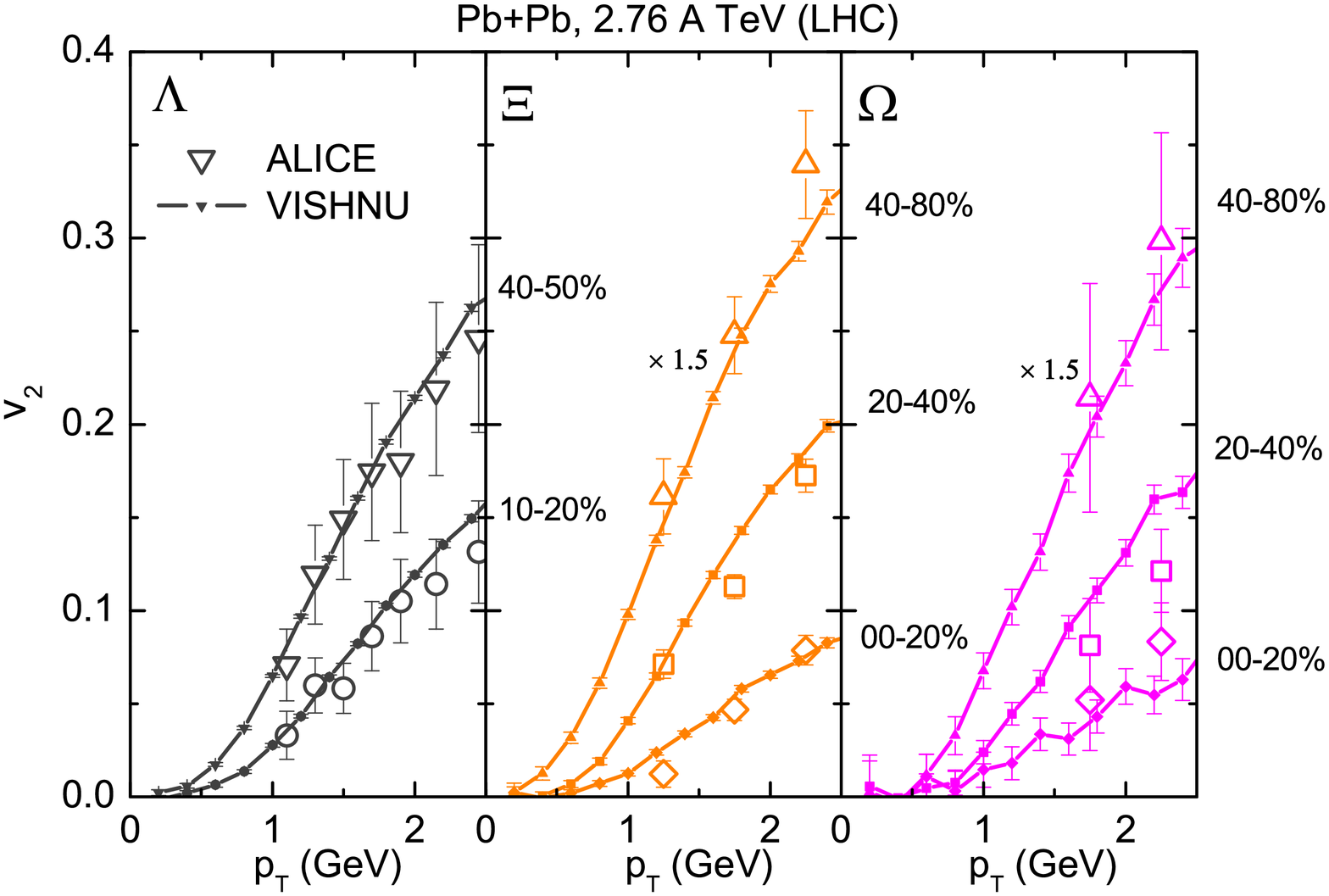}
\caption{(Color online) Differential elliptic flow for $\Lambda$, $\Xi$ and $\Omega$ in 2.76 A TeV Pb+Pb collisions. Experimental data are from ALICE~\cite{Zhou:2013uwa}, theoretical lines are from  {\tt VISHNU} simulations with $(\eta/s)_{QGP}=0.16$ and {\tt MC-KLN} initial conditions.
\label{F4}
}
\end{figure*}
%
\begin{figure}[t]
\hspace{-7mm}
\includegraphics[width=0.40\linewidth,height=7cm,clip=]{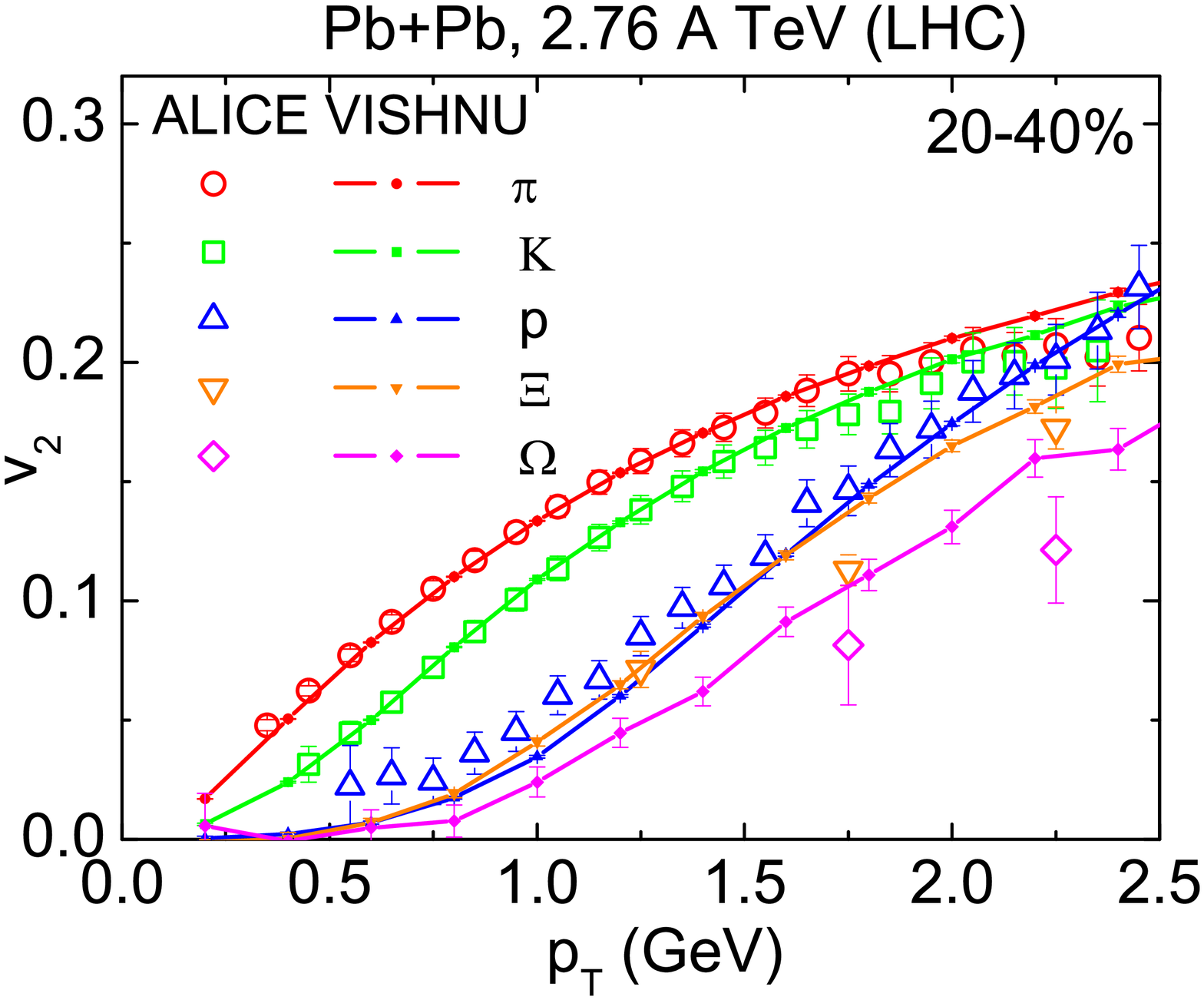}
\hspace{-15mm}
\includegraphics[width=0.40\linewidth,height=7cm,clip=]{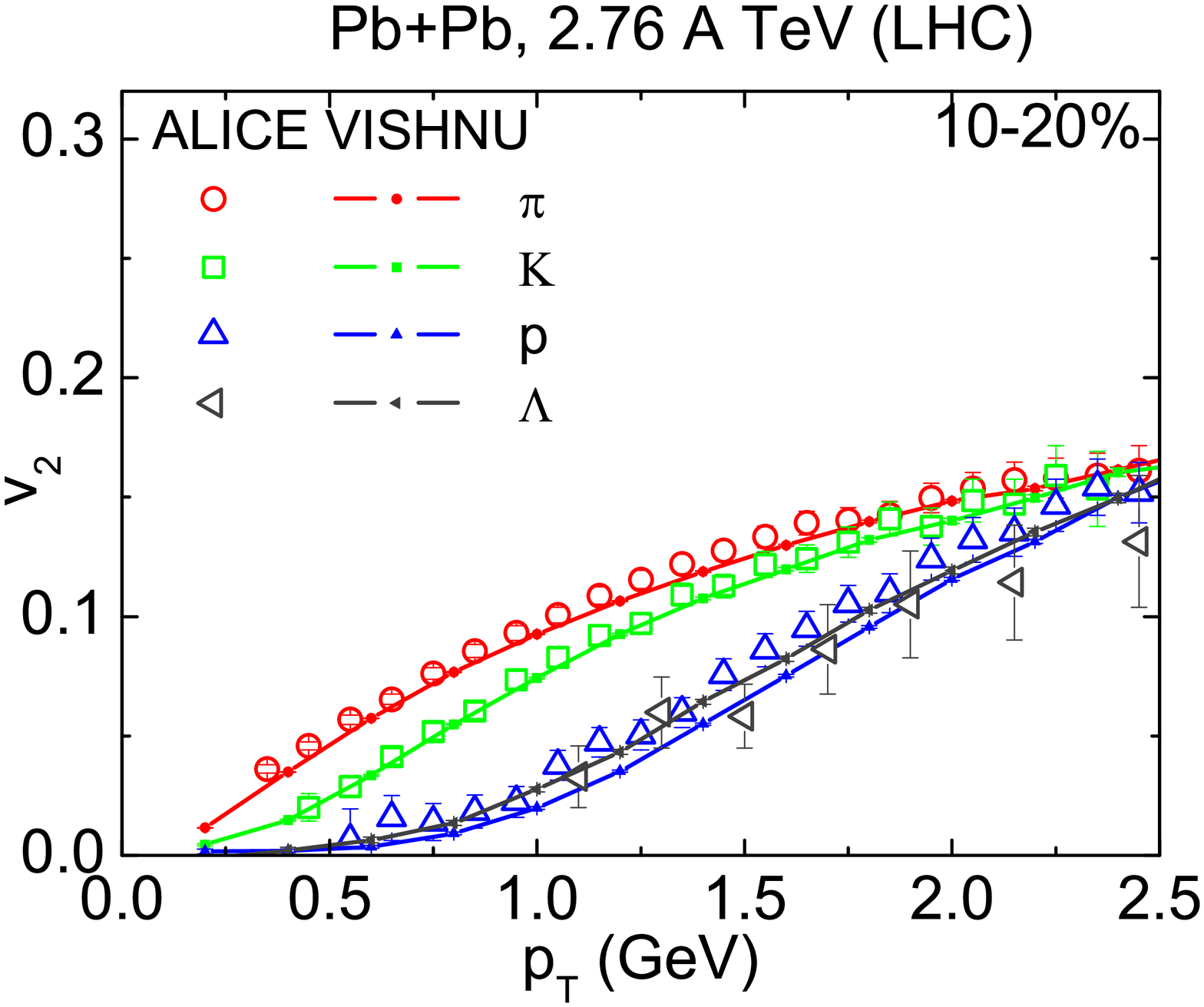}
\hspace{-15mm}
\includegraphics[width=0.40\linewidth,height=7cm,clip=]{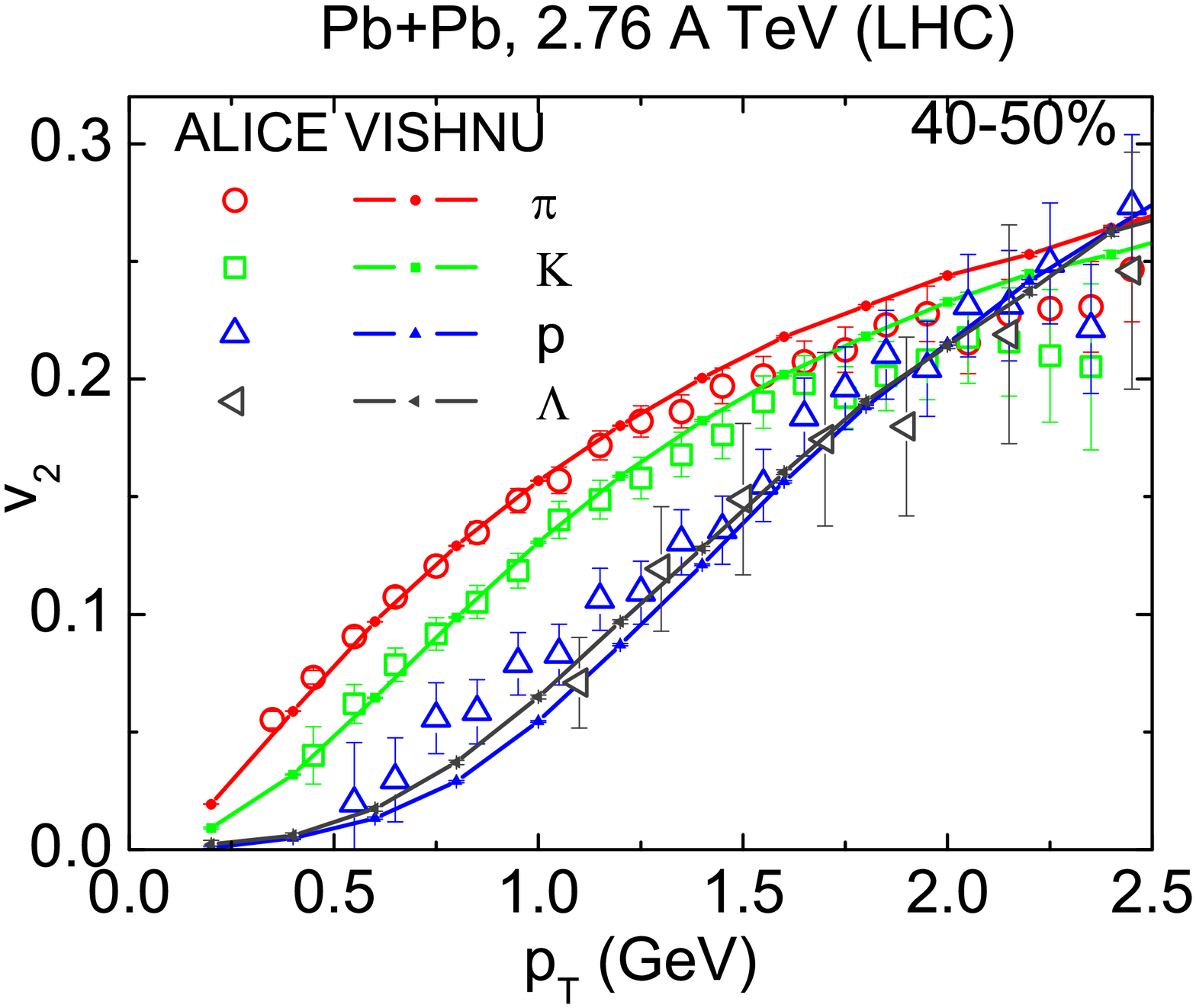}
\caption{(Color online) Mass ordering of elliptic flow for various hadrons in 2.76 A TeV Pb+Pb collisions. The theoretical lines for pions, kaons and protons are from earlier  {\tt VISHNU} calculations in Ref.~\cite{Song2013}.
\label{F4}
}
\end{figure}

\section{Results}
\quad {\tt VISHNU} hybrid model is a successful tool to describe and predict the soft hadron productions at RHIC and the LHC. Using {\tt VISHNU}~\cite{Song:2010aq}, we extracted the QGP specific shear viscosity from the integrated elliptic flow for all charged hadrons in 200 A GeV Au+Au collisions and found $ 1/(4\pi)<(\eta/s)_{QGP}<2.5/(4\pi)$ ($ 0.08 <(\eta/s)_{QGP}<0.20$), where the uncertainties are dominated by the undetermined initial conditions from {\tt MC-Glauber} and {\tt MC-KLN} models~\cite{Song:2010mg}. With that extracted QGP viscosity, {\tt VISHNU} gives a nice description of the $p_T$ spectra and elliptic flow for all charged and identified hadrons at RHIC~\cite{Song:2011hk}.

Ref.~\cite{Song:2011qa} extrapolates the {\tt VISHNU} calculations to the LHC and shows that {\tt VISHNU} could nicely describe the $p_T$ spectra and elliptic flow for all charged hadrons with {MC-KLN} initial conditions and $(\eta/s)_{QGP}=0.20-0.24$, a slightly higher value than the one extracted at RHIC with MC-KLN initial conditions. With the identified soft hadron data becoming available, we further calculated the multiplicity, $p_T$ spectra and differential elliptic flow for pions kaons and protons and found these $v_2(p_T)$ prefer $(\eta/s)_{QGP}=0.16$ for {\tt MC-KLN}~\cite{Song2013,Heinz:2011kt}. It is noticed that the integrated and differential elliptic flow for all charged hadrons are measured by 4 particle cumulants method ($v_2\{4\}$), while the elliptic flow for identified hadrons are measured by scalar product method ($v_2\{sp\}$) based on 2 particle correlations. Comparing with $v_2\{4\}$, non-flow and fluctuations raise the flow signal of $v_2\{sp\}$, leading to a slightly lower value of $(\eta/s)_{QGP}$ to fit $v_2\{sp\}$ data for identified hadrons.

Recently, the ALICE collaboration measured the elliptic flow for $\Lambda$, $\Xi$ and $\Omega$ in 2.76 A TeV Pb+Pb collisions using scalar product method~\cite{Zhou:2013uwa}. Using {\tt VISHNU} hybrid model, we calculate the differential elliptic flow $v_2(p_T)$ for these strange and multi-strange hadrons. Here we choose $(\eta/s)_{QGP}=0.16$, {\tt MC-KLN} initial conditions and other parameter sets as used in early calculations~\cite{Song2013} that nicely describe the soft hadron data for pions kaons and protons at the LHC. Fig.~1 shows that {\tt VISHNU} nicely describes $v_2$($p_T$) data for $\Xi$ and $\Omega$ from central to peripheral collisions and nicely describes $\Lambda$ data at 10\% and 40\% centralities as measured by ALICE. Due to the very low particle yields, the elliptic flow for $\Omega$ has large error bars in ALICE data and the {\tt VISHNU} results. For better comparisons, the statistics for $\Omega$ should be further increased in both experimental and theoretical sides.

Fig.~2 further explores the mass-ordering of elliptic flow among various hadrons in 2.76 A TeV Pb+Pb collisions. Radial flow tends to push low $p_T$ hadrons to higher transverse momenta. Such effects are more efficient for heavier particles, leading to the mass-ordering of $v_2$ as predicted in hydrodynamics~\cite{Huovinen:2001cy}. Comparing with $v_2(p_T)$ for all charge hadrons, the mass-splitting of $v_2$ among various hadrons reflects the interplay of the radial and elliptic flow during hadronic evolution, which is more sensitive to details of theoretical calculations.
Fig.~2 shows that, within current statistics, {\tt VISHNU} roughly reproduces the $v_2$ mass ordering among $\pi$, K, p, $\Xi$ and $\Omega$ in semi-central collisions, but gives an inverse mass-ordering between p and $\Lambda$ for the two selected (10-20\% and 40-50\%) centralities. Early OSU calculations~\cite{Shen:2011eg}, using pure viscous hydrodynamics {\tt VISH2+1}, nicely predicts the mass ordering among $\pi$, K, p, $\Xi$, $\Omega$ and $\Lambda$ as measured in experiments later on, but over-predicts the elliptic flow for proton and $\Lambda$ at 10-20\% centrality bin. With the microscopic description of hadronic rescatterings and evolution, {\tt VISHNU} hybrid model improves the descriptions of the elliptic flow for proton and $\Lambda$, while the slightly under-predicts of the proton $v_2(p_T)$ below 2 GeV lead to an inverse mass ordering of $v_2$ between p and $\Lambda$.  To improve the description of $v_2(p_T)$ for proton as well as for $\Xi$ and $\Omega$, the related hadronic cross sections in {\tt UrQMD} need to be improved. On the other hand, to further understand these rare multi-strange hadrons and to test theoretical models, the statistics for $\Xi$ and $\Omega$ need to be increased in both experimental and theoretical sides. \\[-0.05in]

\section{Summary}
\quad In this proceeding, we present {\tt VISHNU} hybrid model calculations of the elliptic flow for $\Lambda$, $\Xi$ and $\Omega$ in 2.76 A TeV Pb+Pb collisions. {\tt VISHNU} nicely describes $v_2(p_T)$ for these strange and multi-strange hadrons at various centralities measured by ALICE. It roughly reproduce the mass-ordering of $v_2$ among $\pi$, K, p, $\Xi$ and $\Omega$ within current statistics, but gives an inverse mass-ordering between p and $\Lambda$. For a better comparison and understanding of the elliptic flow data for the rare multi-strange hadrons like $\Xi$ and $\Omega$, high statistic runs are needed in both experimental and theoretical sides. \\[-0.05in]

{\it Acknowledgments:}  This work was supported by the new faculty startup funding from Peking University. We gratefully acknowledge extensive computing resources provided by Tianhe-1A at National Supercomputing Center in Tianjin, China.

\smallskip


\section*{References}
\medskip
\begin{thebibliography}{9}
\bibitem{Song:2007fn}
  Song H and Heinz U 2008
  Phys.\ Lett.\  {\bf B658} 279;
  2008 Phys.\ Rev.\  C {\bf 77} 064901;
  2008 Phys.\ Rev.\ C {\bf 78} 024902;
Song H 2009 Ph.D Thesis, The Ohio State University (August 2009),
  arXiv:0908.3656 [nucl-th].

\bibitem{Romatschke:2007mq}
  Romatschke P and Romatschke U 2007
  Phys.\ Rev.\ Lett.\  {\bf 99} 172301;
  Luzum M and Romatschke P 2008
  Phys.\ Rev.\  C {\bf 78} 034915;
  %
  Dusling K and Teaney D 2008
  Phys. Rev. C {\bf 77} 034905;
%
  Molnar D and Huovinen P 2008
  J.\ Phys.\ G {\bf 35} 104125.


\bibitem{Song:2010mg}
  Song H,  Bass S A, Heinz U, Hirano T and Shen C 2011
  Phys.\ Rev.\ Lett.\  {\bf 106} 192301
  [2012 Erratum-ibid.\  {\bf 109} 139904].

\bibitem{Song:2011hk}
  Song H, Bass S A, Heinz U, Hirano T and Shen C 2011
  Phys.\ Rev.\ C {\bf 83} 054910
  [2012 Erratum-ibid.\ C {\bf 86} 059903].


\bibitem{Adams:2003fy}
  Adams J {\it et al.}  [STAR Collaboration] 2004
  Phys.\ Rev.\ Lett.\  {\bf 92} 182301.

\bibitem{Estienne:2004di}
  Estienne M [STAR Collaboration] 2005
  J.\ Phys.\ G G {\bf 31} S873.

\bibitem{Song:2012ua}
 Song H 2013
  Nucl.\ Phys.\ A {\bf 904-905} 114c;
  arXiv:1207.2396 [nucl-th];
  Song H 2014
  arXiv:1401.0079 [nucl-th].

\bibitem{ABELEV:2013zaa}
  Abelev B B {\it et al.}  [ALICE Collaboration] 2013
  arXiv:1307.5543 [nucl-ex];
  Abelev B B {\it et al.}  [ALICE Collaboration] 2013
  arXiv:1307.5530 [nucl-ex].

\bibitem{Zhou:2013uwa}
  Zhou Y [ for the ALICE Collaboration] 2013
  arXiv:1309.3237 [nucl-ex];
  Noferini F [ALICE Collaboration] 2013
  Nucl.\ Phys.\ A904-905 483c.

\bibitem{Song2013-2}
Meng F, Song H, {\it et al.}, in preparation.
%
%
\bibitem{Song2013}
Song H, Bass S A and Heinz U 2013 arXiv:1311.0157 [nucl-th].
%
\bibitem{Song:2010aq}
  Song H, Bass S A and Heinz U 2011
  Phys.\ Rev.\  C {\bf 83} 024912;
  Song H 2012
  Eur.\ Phys.\ J.\ A {\bf 48} 163.



\bibitem{Huovinen:2009yb}
  Huovinen P and Petreczky P 2010
  Nucl.\ Phys.\  {\bf A837} 26.


\bibitem{Song:2011qa}
  Song H, Bass S A and Heinz U 2011
  Phys.\ Rev.\ C {\bf 83} 054912
  [2013 Erratum-ibid.\ C {\bf 87} 019902].


\bibitem{Heinz:2011kt}
  Heinz U, Shen C  and Song H 2012
  AIP Conf.\ Proc.\  {\bf 1441} 766.
%
\bibitem{Huovinen:2001cy}
  Huovinen P, Kolb P F, Heinz U W, Ruuskanen P V and Voloshin S A 2001
  Phys.\ Lett.\ B {\bf 503} 58.
%
%
\bibitem{Shen:2011eg}
  Shen C, Heinz U, Huovinen P and Song H 2011
  Phys.\ Rev.\ C {\bf 84} 044903.







\end{thebibliography}
\end{document}